\documentclass{article}
\usepackage{a4,amsmath,cite}
\usepackage[dvips]{graphicx}

\begin{document}


\title{\textbf{Oscillator model for the relativistic fermion-boson system}}

\author{D.A. Kulikov, 
R.S. Tutik\thanks{tutik@dsu.dp.ua (corresponding author)}\\
\\
{\sl Theoretical Physics Department} \\ {\sl Dniepropetrovsk
National
University }\\
{\sl 72 Gagarina av., Dniepropetrovsk 49010, Ukraine} }

\date{}
\maketitle

\begin{abstract}
{The solvable quantum mechanical model for the relativistic
two-body system composed of spin-$1/2$ and spin-$0$ particles is
constructed. The model includes the oscillator-type interaction
through a combination of Lorentz-vector and -tensor potentials.
The analytical expressions for the wave functions and the order of
the energy levels are discussed.}
\end{abstract}

\bigskip

\section{Introduction}
\label{part1}

Relativistic oscillator models for the spin-1/2 particles are a
subject of rather broad interest nowadays. Apart from the academic
interest in equations with exact solutions, the relativistic
oscillator may serve as the first approximation to description of
the more realistic systems in nuclear and particle physics. In
contrast to the non-relativistic case, within the framework of the
Dirac equation, there exist various possibilities for introducing
the oscillator interaction, such as the mixture of the quadratic
in coordinate vector and scalar potentials with the equal
magnitude and equal or opposite signs
\cite{tassie,tegen,ginocchio}, or the use of the non-minimal
coupling scheme (the so-called Dirac oscillator)
\cite{moshinsky,kukulin}. Note that the Dirac equation with the
oscillator interaction has been applied recently for explaining
the pseudospin symmetry in nuclei \cite{chen,lisboa}.

However, for the careful description of the relativistic effects
in the two-body problem it is needed to go beyond the one-particle
approximation. For this end, the Breit-type equations
\cite{breit,krolikowski,bijtebier,tanaka}, the Barut method
\cite{barut,klimek} and the relativistic quantum mechanics with
constraints \cite{rohrlich,sazdjian86,crater87,crater04} are
usually used. With these methods, the models for the relativistic
two-boson \cite{droz77,rohrlich} and two-fermion
\cite{mesa,nikitin} oscillators have been offered.

Recently, a new approach to the relativistic two-body problem,
based on the extension of the $SL(2,C)$ group to the $Sp(4,C)$
one, has been proposed \cite{kulikov}. This approach has been
applied to construct the wave equation for the two-body system
consisted of the spin-$1/2$ and spin-$0$ particles with
interaction described by means of the Lorentz-vector and
Lorentz-tensor potentials \cite{fb}.

The goal of this work is to find, in addition to the mentioned
two-boson and two-fermion oscillator models, an exact
oscillator-like solution to the obtained equation \cite{fb}
describing the fermion-boson system. In the lack of the
Lorentz-vector interaction, this solution can be regarded as the
generalization of the model of the one-body Dirac oscillator
\cite{moshinsky} to the fermion-boson case.

\section{Wave equation with the extension
of the $SL(2,C)$ group} \label{part2}

In this Section, for completeness of the discussion let us briefly
consider the wave equation for a fermion-boson system derived with
the extension of the $SL(2,C)$ group.

It is well known that the homogeneous Lorentz group $SO(1,3)$ is
covered by the symplectic $Sp(2,C) \equiv SL(2,C)$ group. As a
consequence, the relativistic field theory in the four-dimensional
space-time can equivalently be formulated entirely within the
framework of $Sp(2,C)$ Weyl spinor formalism \cite{penrose}. It
has been shown \cite{kulikov} that the extension of the $Sp(2,C)$
group to the $Sp(4,C)$ one permits us to develop a procedure of
constructing relativistic wave equations for the two-body systems.

Following Ref.~\cite{fb}, we consider the system consisted of a
spin-1/2 fermion and a spin-0 boson which interact with each other
by virtue of the Lorentz-vector and Lorentz-tensor potentials. The
wave function of this system is represented by a Dirac spinor or,
in our treatment, by two $Sp(4,C)$ Weyl spinors $\varphi$ and
$\bar{\chi}$ and the corresponding wave equation has the
Dirac-like form
\begin{eqnarray}\label{eq10}
&&[I\otimes \sigma^m (w_m+A_m)+\tau^1\otimes \sigma^m
(p_m+B_m)]\bar{\chi}                                           \nonumber \\
&&=(m_{+}+\tau^1\otimes I m_{-}
-\mathrm{i}I\otimes\sigma^m\tilde{\sigma}^n C_{mn}
-\mathrm{i}\tau^1\otimes\sigma^m\tilde{\sigma}^n D_{mn})\varphi, \nonumber \\
 &&[I \otimes \tilde{\sigma}^m
(w_m+A_m)+\tau^1\otimes\tilde{\sigma}^m
(p_m+B_m)]\varphi \nonumber \\
&&=(m_{+}+\tau^1\otimes I m_{-}
-\mathrm{i}I\otimes\tilde{\sigma}^m\sigma^n C_{mn}
-\mathrm{i}\tau^1\otimes\tilde{\sigma}^m\sigma^n
D_{mn})\bar{\chi}.
\end{eqnarray}
Here four-momenta $w_m$, $p_m$ and mass parameters $m_{+}$,
$m_{-}$ are related to the quantities for the particles of the
system by
\begin{equation}\label{eq2}
w_m=\frac{1}{2}(p_{1m}+p_{2m}),\qquad
p_m=\frac{1}{2}(p_{1m}-p_{2m}),
\end{equation}
\begin{equation}\label{eq3}
m_{+}=\frac{1}{2}(m_{1}+m_{2}),\qquad
m_{-}=\frac{1}{2}(m_{1}-m_{2}),
\end{equation}
and the Lorentz-vector potentials $A_m$, $B_m$ as well as the
Lorentz-tensor ones $C_{mn}$, $D_{mn}$ in general case are the
functions of $w_m$, $p_m$ and the relative coordinate
$x_m=x_{1m}-x_{2m}$. The direct matrix products contain $\sigma^m$
and $\tilde{\sigma}^m$ that are the $2\times 2$ matrices from the
one-particle Dirac equation in the Weyl notation, for which we use
the conventional representation: $\sigma^m=(I,\,
\boldsymbol\tau)$, $\tilde{\sigma}^m=(I,\, -\boldsymbol\tau)$
where $I$ is the unit $2\times 2$ matrix and
$\boldsymbol\tau=(\tau^1,\tau^2,\tau^3)$ are the Pauli matrices.
Besides, the Minkowski metrics $h^{mn}=\mathrm{diag}(1,-1,-1,-1)$
is accepted.

Note that the first multipliers in the direct products originate
due to the group extension and are essential for obtaining the
two-particle interpretation of the problem. The above equation
(\ref{eq10}) should be supplemented with the subsidiary condition
\begin{equation}\label{eq4}
(w^m p_m -m_{+}m_{-}) \left(\begin{array}{c}
   \varphi \\
    \bar{\chi}
  \end{array}\right)\equiv \frac{1}{4}(p_1^2-p_2^2-m_1^2+m_2^2) \left(\begin{array}{c}
   \varphi \\
    \bar{\chi}
  \end{array}\right)=0,
\end{equation}
which must guarantee that in the case when the interaction is
absent the particles are on the mass shell and Eqs.~(\ref{eq10})
and (\ref{eq4}) are reduced to the free Dirac and Klein-Gordon
equations. These free equations are derived with decomposing the
spinor wave functions into the projections
\begin{equation}\label{eq14}
\varphi_{\pm}=\frac{1}{2}(1\pm\tau^1\otimes I)\varphi, \qquad
\bar{\chi}_{\pm}=\frac{1}{2}(1\pm\tau^1\otimes I)\bar{\chi}
\end{equation}
that are two-component $Sp(2,C)$ Weyl spinors. The projections
labeled by plus (minus) sign describe the system in which the
Dirac particle has mass $m_1$ ($m_2$) and the Klein-Gordon
particle has mass $m_2$ ($m_1$).

Thus, the wave equation (\ref{eq10}) supplemented with the
subsidiary condition (\ref{eq4}) describes two systems composed of
the spin-$1/2$ and spin-$0$ particles, which differ from each
other only in permutation of masses of the particles.

It is to be pointed that in the presence of the interaction the
wave equation (\ref{eq10}) and the subsidiary condition
(\ref{eq4}) must be compatible. The demand of the compatibility
imposes the following restrictions on the shape of the potentials
\begin{eqnarray}\label{eq11}
&&\omega^m\pi_m+\pi_m\omega^m=2w_m p^m, \qquad
\omega^m D_{mn}-D_{mn}\omega^m+ C_{mn}\pi^m - \pi^mC_{mn}=0,  \nonumber \\
&&C_{mk}D^{mn}+D_{mk}C^{mn}=0
\end{eqnarray}
where $\omega_m=w_m+A_m$, $\pi_m=p_m+B_m$. Moreover, the
potentials must depend on the relative coordinate only through its
transverse part
\begin{equation}\label{eq13}
x_{\bot}^m=(h^{mn}-w^m w^n/w^2)x_n
\end{equation}
with respect to the total four-momentum $w_m$, which is conserved
and so can be treated as the eigenvalue rather than the operator.

%
%

It was shown \cite{fb} that in the presence of the only
Lorentz-vector potentials the discussed approach reproduces the
known equation by Kr\'{o}likowski \cite{krolikowski}, which is
also derived by reducing the Bethe-Salpeter equation
\cite{tanaka}, and the equations of the relativistic quantum
mechanics with constraints \cite{sazdjian86,crater87}. In
addition, being important for the study of mesonic atoms
\cite{kelkar,rusetsky}, the problem of the description of the
electromagnetic interaction in the system of a fermion with an
anomalous magnetic moment and a boson was considered with the
quasipotential \cite{austen} and Breit-type \cite{kelkar}
equations. In our approach, the corresponding potentials for this
problem are
\begin{eqnarray}\label{eq16}
A_m&=&\left(\left(1-\frac{2\mathcal{A}}{E}\right)^{1/2}-1\right)w_m, \nonumber \\
B_m&=&\left(\left(1-\frac{2\mathcal{A}}{E}\right)^{-1/2}-1\right)p_m
+\frac{\mathrm{i}}{2E}\left(1-\frac{2\mathcal{A}}{E}\right)^{-3/2}\frac{\partial
\mathcal{A}}{\partial x_{\bot}^m} , \nonumber \\
C_{mn}&=&0, \qquad D_{mn}=\frac{k_1}{4m_1} \left( \frac{\partial
A_n}{\partial x_{\bot}^m}-\frac{\partial A_m}{\partial
x_{\bot}^n}+\frac{\partial B_n}{\partial
x_{\bot}^m}-\frac{\partial B_m}{\partial x_{\bot}^n} \right)
\end{eqnarray}
where $E=2\sqrt{w^2}$ is the total energy,
$\mathcal{A}=-\alpha/\sqrt{-x_{\bot}^2}$ is the Coulomb potential,
$k_1$ and $m_1$ denote the anomalous magnetic moment and the mass
of the fermion. With these potentials inserted, the wave equation
(\ref{eq10}) in the semirelativistic approximation coincides with
that reported in Ref.~\cite{austen}.

Thus, the equation (\ref{eq10}), derived with the extension of the
$SL(2,C)$ group to the $Sp(4,C)$ one, together with the subsidiary
condition (\ref{eq4}) restores the results obtained in other
approaches. However, including the Lorentz-tensor potentials in
our equation, in addition to the Lorentz-vector ones, permits us
to treat more wide range of problems concerning fermion-boson
interactions and to construct the new oscillator model involved
potentials of both these types of the Lorentz structure.

\section{Exact oscillator-like solution}
\label{part3}

Now we will construct an oscillator model for the relativistic
two-body system composed of the spin-$1/2$ and spin-$0$ particles,
the word ``oscillator" being taken to include all equations that
can be transformed into the second-order equation containing
quadratic in relative coordinate potentials. Such an
oscillator-like second-order equation may arise from the
first-order wave equation (\ref{eq10}) if both the Lorentz-vector
and -tensor potentials in Eq.~(\ref{eq10}) are linear in the
relative coordinate.

However, in the case of the Lorentz-vector potentials, the purely
linear in $x^m_{\bot}$ term can be removed by including this term
into the phase factor of the wave function. Therefore, we suppose
that the Lorentz-vector potentials involve not only the relative
coordinate but also the relative four-momentum. A combination of
these two quantities, having the structure of the multiplication
of the orbital momentum with $x^m_{\bot}$, will give rise to the
desired oscillator interaction. In order to avoid the Klein
paradox, this Lorentz-vector oscillator potential must have the
space-like rather than the time-like structure.

For constructing a Lorentz-tensor oscillator potential, we combine
the linear in the relative coordinate term with the total
four-momentum, which is the constant of motion, and set the
potentials in Eq.~(\ref{eq10}) as follows
\begin{equation}\label{eq22}
\begin{array}{ll}
A_m=0, \qquad &B_m=\lambda (x_{\bot}^2 p_{\bot m}-x_{\bot m}
x_{\bot}^n p_{\bot n}-\mathrm{i}x_{\bot m}),\\
C_{mn}=0, \qquad &D_{mn}=\nu(w_m x_{\bot n}-w_n x_{\bot
m})/2\sqrt{w^2}
\end{array}
\end{equation}
where $\lambda$ and $\nu$ are constants with the dimensionality of
the string tension, and the term with imaginary unit is introduced
for the Hermiticity of the potential. Note that the oscillator
potentials $A_m$ and $C_{mn}$ must be zero because, from
Eqs.~(\ref{eq10})--(\ref{eq3}), they are transformed under the
permutation of the particles like $w_m$ ($w_m\rightarrow w_m$,
$A_m\rightarrow A_m$, $C_{mn}\rightarrow C_{mn}$), but being
linear in $x^m_{\bot}$, as the oscillator potentials, they must be
transformed with the opposite sign.

The wave equation (\ref{eq10}) with such an interaction proved to
be exactly solvable. Since this equation describes two
fermion-boson systems which differ from each other only in
permutation of masses of the particles, we accept that the system
consists of a Dirac fermion with the mass $m_1$ and a Klein-Gordon
boson with the mass $m_2$. To find the solution, we pass to the
spinor projections $\varphi_{+}$ and $\bar{\chi}_{+}$, for which
the wave equation (\ref{eq10}) and the subsidiary condition
(\ref{eq4}), being written in the center-of-mass frame
($\mathbf{w}=0$), are reduced to the Dirac-like equation
\begin{eqnarray}\label{eq24}
&&\left(\frac{E}{2}+\frac{m_1^2-m_2^2}{2E}-m_1\right)\phi=\boldsymbol\tau\cdot(\boldsymbol\pi
+\mathrm{i}\nu\mathbf{x}) \psi, \nonumber \\
&&\left(\frac{E}{2}+\frac{m_1^2-m_2^2}{2E}+m_1\right)\psi=\boldsymbol\tau\cdot(\boldsymbol\pi
-\mathrm{i}\nu\mathbf{x}) \phi
\end{eqnarray}
where $\phi=\bar{\chi}_{+}+\varphi_{+}$,
$\psi=\bar{\chi}_{+}-\varphi_{+}$,
$\boldsymbol\pi=\mathbf{p}+\lambda
(\mathbf{x}\times\mathbf{l}-\mathrm{i}\mathbf{x})$ and
$\mathbf{l}=\mathbf{x}\times\mathbf{p}$ is the orbital momentum of
the relative motion.

In the last equation, the total angular momentum of the relative
motion, $\mathbf{j}=\mathbf{x}\times\mathbf{p}+\boldsymbol\tau/2$,
is conserved. Then the spatial variables are separated and we can
write
\begin{equation}\label{eq25}
\left(\begin{array}{cc}
   \phi \\
   \psi
  \end{array} \right)=\left(
  \begin{array}{cc}
   \frac{G(r)}{r}\mathcal{Y}_{\kappa m}(\Omega) \\
   \frac{\mathrm{i}F(r)}{r}\mathcal{Y}_{-\kappa m}(\Omega)
  \end{array}\right)
\end{equation}
where $\mathcal{Y}_{\kappa m}(\Omega)$ are the so-called spinor
spherical harmonics \cite{greiner} and the normalization of Dirac
spinors imply that for the upper and lower radial functions we
must have
\begin{equation}\label{eq28}
\int_0^{\infty}(G^2(r)+F^2(r))dr=1,
\end{equation}
so that $G(r)$ and $F(r)$ should be square-integrable functions.

The quantum number $\kappa$ is related to the orbital quantum
number $l$ by the expression
\begin{equation}\label{eq26}
\kappa =\left\{
\begin{array}{ccc}
l & =+\left( j+1/2\right) , & l=j+1/2\ \ \ (\kappa>0)\; \\[0.2cm]
-(l+1) & =-\left( j+1/2\right) , & l=j-1/2\ \ \ (\kappa<0)\;
\end{array}
\right.
\end{equation}
that completely determines the orbital and total angular momenta
\begin{equation}\label{eq27}
l=|\kappa|+\frac{1}{2}\left(\frac{\kappa}{|\kappa|}-1\right),
\qquad j=|\kappa|-\frac{1}{2}
\end{equation}
and, hence, the parity $(-1)^l$.

Inserting (\ref{eq25}) into (\ref{eq24}), we obtain the set of
radial equations
\begin{eqnarray}\label{eq29}
&&\left(\frac{\mathrm{d}}{\mathrm{d}r}+\frac{\kappa}{r}+(\nu-\lambda\kappa)
r\right )G(r)-
\left(\frac{E}{2}+\frac{m_1^2-m_2^2}{2E}+m_1\right)F(r)=0,
 \\
&&\left(\frac{\mathrm{d}}{\mathrm{d}r}-\frac{\kappa}{r}-(\nu-\lambda\kappa)
r\right )F(r)+
\left(\frac{E}{2}+\frac{m_1^2-m_2^2}{2E}-m_1\right)G(r)=0
\nonumber
\end{eqnarray}
which, with eliminating $F(r)$, can be converted into the
oscillator-type equation
\begin{equation}\label{eq30}
\left(\frac{\mathrm{d}^2}{\mathrm{d}r^2}-\frac{\kappa(\kappa+1)}{r^2}
-(\nu-\lambda\kappa)^2
r^2-(\nu-\lambda\kappa)(2\kappa-1)+\mathcal{E}\right)G(r)=0
\end{equation}
where
\begin{equation}\label{eq31}
\mathcal{E}=\frac{E^2}{4}-\frac{m_1^2+m_2^2}{2}
+\frac{(m_1^2-m_2^2)^2}{4E^2}.
\end{equation}

This equation possesses the exact solution that reads as
\begin{equation}\label{eq32}
G(r)=A\exp(-\frac{1}{2}a^2 r^2)(a^2 r^2)^{(l+1)/2}L_n^{l+1/2}(a^2
r^2)
\end{equation}
where $a=\sqrt{|\nu-\lambda\kappa|}$, $n=0,1,...$ denotes the
radial quantum number and $A$ is a normalization constant.

The lower radial function is obtained from Eqs.~(\ref{eq29}) with
using the recursion relations for the generalized Laguerre
polynomials \cite{stegun}.

For $\kappa>0$ we have
\begin{eqnarray}\label{eq33}
F(r)=&&\frac{aEA}{(E+m_1)^2-m_2^2}\exp(-\frac{1}{2}a^2 r^2)(a^2
r^2)^{l/2}\times\nonumber \\
&&\left[ (1+s)(n+l+1/2)L_{n}^{l-1/2}(a^2
r^2)+(1-s)(n+1)L_{n+1}^{l-1/2}(a^2 r^2)\right]
\end{eqnarray}
and, for $\kappa<0$,
\begin{eqnarray}\label{eq34}
F(r)=&&\frac{aEA}{(E+m_1)^2-m_2^2}\exp(-\frac{1}{2}a^2 r^2)(a^2
r^2)^{(l+2)/2}\times\nonumber \\
&&\left[ (1+s)L_{n}^{l+1/2}(a^2 r^2)-2L_{n}^{l+3/2}(a^2
r^2)\right]
\end{eqnarray}
where $s=\mathrm{sgn}(\nu-\lambda\kappa)$.

Then the energy eigenvalues of Eq.~(\ref{eq30}) are defined by
\begin{equation}\label{eq37}
\mathcal{E}=|\nu-\lambda\kappa|[4n+2l+3+\mathrm{sgn}(\nu-\lambda\kappa)(2\kappa-1)].
\end{equation}
Hence
\begin{equation}\label{eq36}
E^2=m_1^2+m_2^2+2\mathcal{E}+2\sqrt{(m_1^2+\mathcal{E})(m_2^2+\mathcal{E})}
\end{equation}
describes the energy levels with quantum numbers $(n,\kappa )$ or,
using the standard spectroscopic notation, the energies of the
states $n\,l_{j}$, because $\kappa$ determines uniquely $l$ and
$j$.

It should be pointed that the appearance of the signum function in
Eq.~(\ref{eq37}) imposes a restriction on the permitted values of
the coupling constants $\lambda$ and $\nu$. For example, if $\nu >
\lambda>0$, there always exists a value of $l$ such that for
$\kappa=l$ we have $\mathrm{sgn}(\nu-\lambda\kappa)=+1$ whereas
for $\kappa'=l'=l+1$ we get $\mathrm{sgn}(\nu-\lambda\kappa')=-1$.
Then from Eq.~(\ref{eq37}) it follows that for $l\leq
\nu/\lambda-1/2$ the level $n\,(l+1)_{j+1}$ will be lower than the
level $n\,l_{j}$. For avoiding this unphysical order of levels, it
is sufficient to demand that either $|\lambda|\geq |\nu|$ or
$\lambda=0$. In the following, only the values of $\lambda$ and
$\nu$ which obey these conditions are considered.

\section{Discussion} \label{part4}

The derived energy spectrum of the fermion-boson system with the
discussed  oscillator interaction has proved to be essentially
distinctive from that of the non-relativistic harmonic oscillator.
The reason is that both the Lorentz-vector and -tensor potentials
contribute to the strong spin-orbit coupling. In the expression
(\ref{eq37}) for the eigenenergies, the spin-orbit coupling is
described by the term containing $(2\kappa-1)$ [note that
$\boldsymbol\tau\cdot\mathbf{l}=-(\kappa+1)$]. This term breaks
the $(2n+l)$-degeneracy, which is inherent in the spectrum of
non-relativistic harmonic oscillator. Moreover, in
Eq.~(\ref{eq37}) a common factor $|\nu-\lambda\kappa|$, playing
the role of an effective oscillator frequency, is
$\kappa$-dependent, too. This implies that the spectrum is not
equidistant except for the case when the Lorentz-vector
interaction is absent ($\lambda = 0$).

\begin{figure}[!ht]
\begin{center}
\includegraphics[width=9cm]{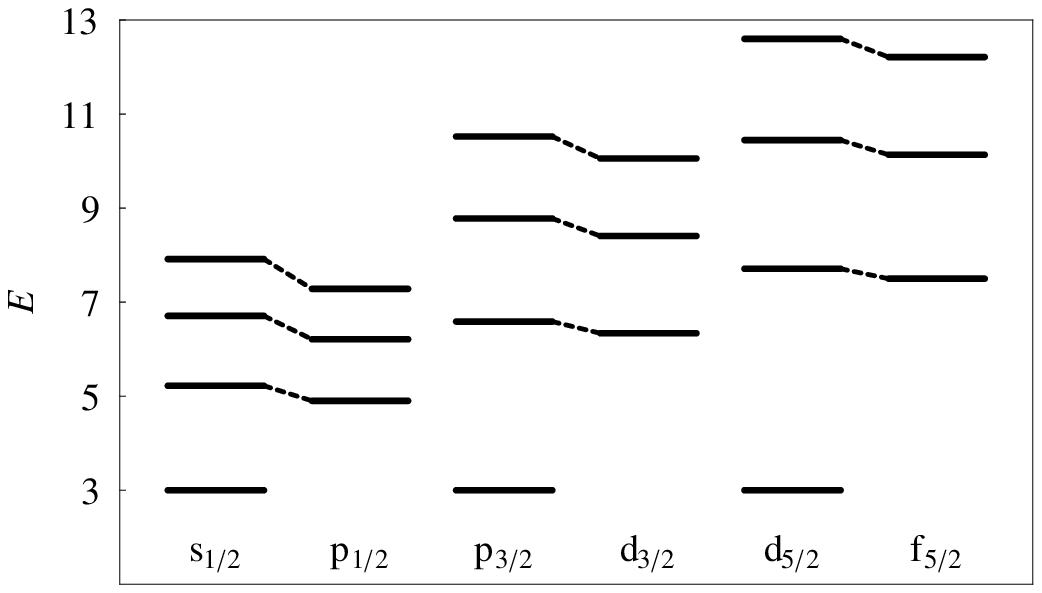}
\end{center}
\par
\vspace*{-0.4cm} \caption{Energy levels for  $\lambda=1$,
$\nu=0.1$, $m_1=1$ and $m_2=2$ in natural units. The levels with
the same $j$ which become degenerate when $\nu=0$ are connected by
dashed lines.} \label{Fig:spek1}
\end{figure}

\begin{figure}[!ht]
\begin{center}
\includegraphics[width=9cm]{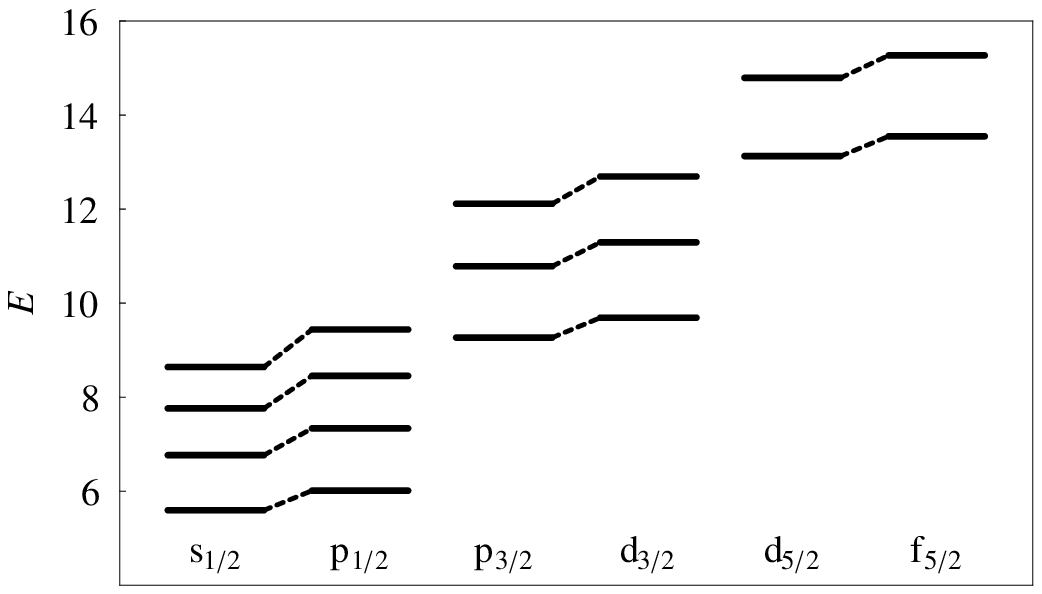}
\end{center}
\par
\vspace*{-0.4cm} \caption{Energy levels for  $\lambda=-1$,
$\nu=0.1$, $m_1=1$ and $m_2=2$ in natural units. The levels with
the same $j$ which become degenerate when $\nu=0$ are connected by
dashed lines.} \label{Fig:spek2}
\end{figure}

If $\lambda \neq 0$, the spacing of the energy levels depends on
whether $\lambda$ is positive or negative. For the case
$\lambda>0$, a typical spacing of the first energy levels is shown
in Fig.~{\ref{Fig:spek1}} where the spectrum was evaluated with
parameters $\lambda=1$, $\nu=0.1$, $m_1=1$ and $m_2=2$ in natural
units. From Fig.~{\ref{Fig:spek1}} it can be seen that the
ground-state levels with $n=0$ and $l=j-1/2=0,1,..$ are degenerate
and correspond to the lowest energy eigenvalue $E=m_1+m_2$. In
contrast to this, the degeneracy does not occur in the case
$\lambda<0$, as seen from Fig.~{\ref{Fig:spek2}}, in which the
energy levels are evaluated using $\lambda=-1$, $\nu=0.1$, $m_1=1$
and $m_2=2$.

To proceed, let us consider a particular case when $\lambda \neq
0$ and $\nu=0$, i.e. when the only Lorentz-vector interaction is
present. It should be noticed that in this case our model reduces
to that offered within the framework of the relativistic quantum
mechanics with constraints \cite{sazdjian86}. In
Ref.~\cite{sazdjian86}, the subcase $\lambda>0$ was considered and
the corresponding energy spectrum was proved to display the
parity-doubling phenomenon. However, we have found that in the
other subcase ($\lambda<0$) not only the exact oscillator-like
solution exists, but also the parity doubling has a different
meaning.

For $\lambda>0$ the parity doubling means that the state with
quantum numbers $(n,j,l=j+1/2)$ is degenerate with the state with
the quantum numbers $(n+1,j,l=j-1/2)$, as can be checked by
putting $\nu=0$ in Eq.~(\ref{eq37}) [see also
Fig.~{\ref{Fig:spek1}}]. On the contrary, for $\lambda<0$ the
parity doubling reveals itself as the degeneracy of the states
with $(n,j,l=j+1/2)$ and $(n,j,l=j-1/2)$ which correspond to the
same value of $n$ [see Fig.~{\ref{Fig:spek2}}].

A few remarks concerning the parity doubling are in order. There
exists reasonably strong experimental evidence that orbitally
excited baryons occur in pairs of nearly degenerate states of
opposite parity (for a review see \cite{jaffe}). In our oscillator
model, the parity doubling of this kind takes place in the subcase
$\lambda<0$, whereas for $\lambda>0$ the spectrum includes the
degenerate ground states with $n=0$ and $l=j-1/2$ which are absent
in the baryon spectra. In addition, the appearance of the parity
doubling is closely related to the Lorentz-structure of the
involved potentials. So, in the generalized Nambu-Jona-Lasinio
model of hadrons the parity doubling  occurs when the space-like
Lorentz-vector potential dominates over the Lorentz-scalar one
\cite{kalashnikova,glozman}. Analogously, in our approach the
parity doubling appears when $\nu=0$ and the potential becomes a
pure spatial Lorentz vector.

Now let us turn to the second particular case, in which
$\lambda=0$, i.e., the pure Lorentz-tensor interaction is
retained. In this case our oscillator-like solution can be treated
as the generalization of the  one-body Dirac oscillator model
\cite{moshinsky} to the fermion-boson case. Recall that the Dirac
oscillator is obtained by replacing the momentum $\mathbf{p}$ in
the Dirac equation by
\begin{equation}\label{eq60}
\mathbf{p}\rightarrow \mathbf{p}-\mathrm{i}m \omega\beta \mathbf{x}
\end{equation}
where $\beta$ is the usual Dirac matrix and $\omega$ is the
oscillator frequency.  It is obvious that Eq.~(\ref{eq24}) for the
fermion-boson system transforms into the equation for the Dirac
oscillator after inserting $\lambda=0$, $\nu=m_1 \omega$ and
passing to the one-particle limit, in which the boson has a mass
much larger than the fermion. Note that the passage to this
one-particle limit does not affect the interaction terms in
Eq.~(\ref{eq24}). Therefore, in the case $\lambda=0$ the spectrum
of the fermion-boson system will have the same order of the energy
levels as the spectrum of the Dirac oscillator.

In conclusion, we have constructed the relativistic oscillator
model for the two-body system consisted of the spin-1/2 fermion
and the spin-0 boson interacting by virtue of the Lorentz-vector
and Lorentz-tensor potentials. For this model, the analytical
expressions for the wave functions and the energy eigenvalues have
been obtained. The exact solubility of the oscillator model
suggests it may be used as the first approximation in the study of
more complicated systems in hadron and nuclear physics.


\section*{Acknowledgments}

We thank Prof. T. Tanaka for drawing our attention to Ref.~
\cite{tanaka} and discussing its details. This research was
supported by a grant N 0106U000782 from the Ministry of Education
and Science of Ukraine which is gratefully acknowledged.


\end{document}